\documentstyle[12pt]{article}

\begin{document}

\begin{titlepage}

\begin{center}
{\Large\bf Scheme Dependence in Polarized Deep Inelastic Scattering}
\end{center}
\vskip 2cm

\begin{center}
{\bf Elliot Leader}\\
{\it Department of Physics\\
Birkbeck College, University of London\\
Malet Street, London WC1E 7HX, England\\
E-mail: e.leader@physics.bbk.ac.uk}\\
\vskip 0.5cm
{\bf Aleksander V. Sidorov}\\
{\it Bogoliubov Theoretical Laboratory\\
 Joint Institute for Nuclear Research\\
141980 Dubna, Russia\\
 E-mail: sidorov@thsun1.jinr.ru}
\vskip 0.5cm
{\bf Dimiter B. Stamenov \\
{\it Institute for Nuclear Research and Nuclear Energy\\
Bulgarian Academy of Sciences\\
Blvd. Tsarigradsko Chaussee 72, Sofia 1784, Bulgaria\\
E-mail:stamenov@inrne.bas .bg }}
\end{center}
 
\vskip 0.3cm
\begin{abstract}
We study the effect of scheme dependence upon the NLO QCD 
analysis of the world data on polarized DIS. The
reliability of an analysis at NLO is demonstrated by the
consistency of our polarized densities with the NLO
transformation rules relating them to each other. We stress the
importance of the chiral JET scheme in which all the hard effects
are consistently absorbed in the Wilson coefficient functions.\\

PACS numbers:13.60.Hb; 13.88+e; 14.20.Dh; 12.38.-t
\end{abstract}
\vskip 0.5 cm

\end{titlepage}

\newpage

\setcounter{page}{1}

\vskip 4mm

There has been a major effort in the past decade to obtain
reliable information about the polarized parton densities in the
nucleon, and especially to try to determine the degree of
polarization of the gluon. Aside from its fundamental interest
and its potential as a testing ground for QCD, such information
is vital for the planning of experiments at the RHIC collider,
due to come into operation in 1999.\\

A great improvement in the quality of the data on inclusive deep 
inelastic scattering of leptons on nucleons has been achieved 
recently \cite{E154, newSMCpQ2, finSLACpd} and, concomitantly, 
several detailed NLO QCD analyses of the data have been carried 
out \cite{GSt} - \cite{LSiSt98}, of which the most comprehensive 
are in refs. \cite{Alta, LSiSt98}.\\

\def\thefootnote{\dagger}
It is well known that at NLO and beyond, parton densities become
dependent on the renormalization ( or factorization ) scheme
employed.\footnote{Of course, physical quantities such as the 
virtual photon-nucleon asymmetry $A_1(x,Q^2)$ and the polarized
structure function $g_1(x,Q^2)$ are independent of choice of the
factorization convention.} It is perhaps less well known that there are
significant differences in the polarized case, as a consequence
of the axial anomaly and of the ambiguity in the handling of
$\gamma_5$ in $n$ dimensions.\\

In this paper we study the effect of carrying out the data
analysis in different schemes, in the $\overline{MS}$, AB and what
is called the JET scheme, whose importance we wish to stress. In
the latter all hard effects are consistently absorbed into the Wilson
coefficient functions. The parton densities in each scheme are,
by definition, related to each other by certain NLO transformation
rules. On the other hand, the $Q^2$ evolution in each scheme is
controlled by the splitting functions (or the anomalous dimensions
in Mellin n-moment space) relevant to that scheme.
Thus, in principle, for any two schemes labelled 1 and 2 and for
any generic polarized density $\Delta f$ one has symbolically:
\begin{equation}
NLO~data~analysis~in~Scheme~1~~~\Rightarrow~~~\Delta f_1~,
\label{f1}
\end{equation}
\begin{equation}
NLO~data~analysis~in~Scheme~2~~~\Rightarrow~~~\Delta f_2~,
\label{f2}
\end{equation}
\begin{equation}
NLO~transformation~rules~on~\Delta f_1~\Rightarrow~~~\Delta f_2
\label{f12}
\end{equation}         
and it is a major test of the stability of the analysis and of
the consistency of the theory that the results for $\Delta f_2$ in
(\ref {f2}) and (\ref {f12}) should coincide. In fact, built into 
the theoretical
structure is the feature that the two results actually should
differ by terms of NNLO order. Hence the degree to which the
results agree is a measure of the reliability of carrying out an
analysis at NLO. Moreover, in the set of schemes we study, the non-
singlet and gluon densities are the same in all the schemes. That this
feature should emerge from the data analysis provides a further
test of the stability and reliability of the analysis. The
importance of an analysis of this kind was first stressed in ref.
\cite{anomdg}.\\

In the unpolarized case the most commonly used schemes are the
$\overline{MS}$, MS and DIS and parton densities in 
different schemes differ from each other by terms of order
$\alpha_s(Q^2)~$, which goes to zero as $Q^2$ increases.

There are two significant differences in the polarized case:

{\it ~i)} The singlet densities $\Delta \Sigma(x, Q^2)$, in two 
different schemes, will differ by terms of order
\begin{equation}
\alpha_s(Q^2)\Delta G(x,Q^2)~,
\label{aldelG}
\end{equation}
which appear to be of order $\alpha_s$. But it is known 
\cite{Efr, CaCoMu} that, as a consequence of the axial anomaly, 
the first moment of the polarized gluon density $\Delta G(x,Q^2)$
\begin{equation}
\int _{0}^{1} dx\Delta G(x,Q^2) \propto [\alpha_s(Q^2)]^{-1}~,
\label{delG1}
\end{equation}
and thus grows in such a way with $Q^2$ as to compensate for the 
factor $\alpha_s(Q^2)$ in (\ref{aldelG}). Thus the difference between 
$\Delta \Sigma$ in different schemes is only apparently of order 
$\alpha_s(Q^2)$, and could be quite large.

{\it ii)} Because of ambiguities in handling the renormalization
of operators involving $\gamma_5$ in $n$ dimensions, the specification
$\overline{MS}$ does {\it not} define a unique scheme. Really
there is a family of $\overline{MS}$ schemes which, strictly,
should carry a sub-label indicating how $\gamma_5$ is handled. What
is now conventionally called $\overline{MS}$ is in fact the
scheme due to Vogelsang and Mertig and van Neervenen \cite
{nlocor}, in which the first moment of the non-singlet densities
is conserved, i.e. is independent of $Q^2$, corresponding to the
conservation of the non-singlet axial-vector Cabibbo currents.

Although mathematically correct it is a peculiarity of this
factorization scheme that certain soft contributions are included
in the Wilson coefficient functions, rather than being absorbed
completely into the parton densities. As a consequence, the first
moment of $\Delta \Sigma$ is not conserved so that it is difficult
to know how to compare the DIS results on $\Delta \Sigma$ with
the results from constituent quark models at low $Q^2$.

To avoid these idiosyncrasies Ball, Forte and Ridolfi
\cite{anomdg} introduced what they called the AB scheme, which
involves a minimal modification of the $\overline{MS}$ scheme,
and for which the transformation equations are:
\begin{eqnarray}
\nonumber
\Delta \Sigma(x,Q^2)_{AB}&=&\Delta \Sigma(x,Q^2)_{\overline{MS}}+
N_f{\alpha_s(Q^2)\over 2\pi}
\int _{x}^{1} {dy\over y}\Delta G(y,Q^2)_{\overline{MS}}~,\\
\Delta G(x,Q^2)_{AB}&=&\Delta G(x,Q^2)_{\overline{MS}}
\label{ABMS}
\end{eqnarray}
or, in Mellin $n$-moment space, 
\begin{eqnarray}
\nonumber
\Delta \Sigma(n,Q^2)_{AB}&=&\Delta \Sigma(n,Q^2)_{\overline{MS}}+
N_f{\alpha_s(Q^2)\over 2\pi n}
\Delta G(n,Q^2)_{\overline{MS}}~,\\
\Delta G(n,Q^2)_{AB}&=&\Delta G(n,Q^2)_{\overline{MS}}~.
\label{ABMSmom}
\end{eqnarray}

That $\Delta \Sigma(n=1)_{AB}$ is independent of $Q^2$ to all orders
follows from the Adler-Bardeen theorem \cite{AdlBard}. In (\ref{ABMS}) 
and (\ref{ABMSmom}) $N_f$ denotes the number of flavours.

The singlet part of the first moment of the structure function $g_1$
\begin{equation}
\Gamma^{(s)}_1(Q^2)\equiv \int _{0}^{1} dxg_1^{(s)}(x,Q^2)
\label{Gam1s}
\end{equation}
then depends on $\Delta \Sigma$ and $\Delta G$ only in the
combination 
\begin{equation}
a_0(Q^2)=\Delta \Sigma(1,Q^2)_{\overline{MS}}=
\Delta \Sigma(1)_{AB}-N_f{\alpha_s(Q^2)\over 2\pi}
\Delta G(1,Q^2)
\label{a0}
\end{equation}
and the unexpectedly small value for the axial charge $a_0$ found 
by the EMC
\cite{EMC}, which triggered the "spin crisis in the parton model"
\cite{ELA}, can be nicely explained as due to a cancellation
between a reasonably sized $\Delta \Sigma(1)$ and the gluon
contribution. Of importance for such an explanation are
both the positive sign and the large value (of order 
${\cal O}(1)$) for the first moment of the polarized gluon density 
$\Delta G(1,Q^2)$ at {\it small} $Q^2\sim 1-10~GeV^2$. Note that
what follows from QCD is that $\vert {\Delta G(1,Q^2)}\vert$ 
grows with $Q^2$ (see Eq. (\ref {delG1})) but its value at 
{\it small} $Q^2$ is unknown in the theory at present and has to 
be determined from experiment.\\ 
  
Although the AB scheme corrects the most glaring weakness of the 
$\overline{MS}$ scheme, it does not consistently put all hard
effects into the coefficient functions. As pointed out in 
\cite{Zijlstra}
one can define a family of schemes labelled by a parameter $a$:
\begin{equation}
\pmatrix {\Delta \Sigma \cr \Delta G}_{a}=     			 
\pmatrix {\Delta \Sigma \cr \Delta G}_{\overline{MS}} 
+{\alpha_{s}\over 2\pi}\pmatrix{0  &z(a)_{qG} \cr
                                0  &0}\otimes
\pmatrix {\Delta \Sigma \cr \Delta G}_{\overline{MS}}
\label{afamily}
\end{equation}
where
\begin{equation}
z_{qG}(x;a) = N_f[(2x-1)(a-1)+2(1-x)]~,
\label{zqG}
\end{equation}
in all of which (\ref{a0}) holds, but which differ in their
expression for the higher moments. (The AB scheme corresponds to
taking $a=2$).

Amongst these we believe there are compelling reasons to choose
what we shall call the JET scheme ($a=1$), i.e.
\begin{equation}
z^{JET}_{qG} = 2N_f(1-x)~.
\label{zjet}
\end{equation}

\def\thefootnote{\dagger}
This is the scheme originally suggested by Carlitz, Collins and
Mueller \cite{CaCoMu} and also advocated by Anselmino, Efremov,
Leader and Teryaev in refs. \cite{EfrTer, AEL}.\footnote{There 
is misprint in Eq. (8.2.6) of \cite{AEL}. 
The term $ln({1-x/x^{\prime}\over x/x^{\prime}})$ should be
$[ln({1-x/x^{\prime}\over x/x^{\prime}})-1]$.} In it all hard 
effects are absorbed into the coefficient functions. In this scheme 
the gluon coefficient function is exactly the one that would appear 
in the cross section for 
\begin{equation}
pp\rightarrow jet({\bf k}_{T}) + jet(-{\bf k}_{T})+X~,
\label{2jetprod}
\end{equation}
i.e., the production of two jets with large transverse momentum
${\bf k}_{T}$ and $-{\bf k}_{T}$, respectively.

More recently M\"{u}ller and Teryaev \cite{MulTer} have advanced
rigorous and compelling arguments, based upon a generalization of
the axial anomaly to bilocal operators, that removal of all
anomaly effects from the quark densities leads to the JET scheme.
Also a different argument by Cheng \cite{Cheng} leads to the same
conclusion. (Cheng calls the JET scheme a chirally invariant (CI)
scheme.)

The transformation from the $\overline{MS}$ scheme of Mertig, van
Neerven and Vogeslang to the JET scheme is given in moment space
by  
\begin{eqnarray}
\nonumber
\Delta \Sigma(n,Q^2)_{JET}&=&\Delta \Sigma(n,Q^2)_{\overline{MS}}+
2N_f{\alpha_s(Q^2)\over 2\pi n(n+1)}
\Delta G(n,Q^2)_{\overline{MS}}~,\\
\Delta G(n,Q^2)_{JET}&=&\Delta G(n,Q^2)_{\overline{MS}}~.
\label{JETMS}
\end{eqnarray}

Note that (\ref{JETMS}) implies that the strange sea $\Delta\bar{s}$ 
is different in the two schemes. Of course, (\ref{ABMSmom}) and 
(\ref{JETMS}) become the same for $n=1$.\\

\def\thefootnote{\dagger}
The NLO Wilson coefficient functions $\Delta C^{(1)}_i(x)$ and 
polarized splitting functions $\Delta P^{(1)}_{ij}(x)$ (or the
corresponding anomalous dimensions $\Delta \gamma^{(1)}_{ij}(n)$) for 
the $\overline{MS}$ and AB schemes can be found in refs. \cite{nlocor} and
\cite{anomdg}, respectively. The NLO coefficient functions and
anomalous dimensions in the JET scheme are related to those of the
$\overline{MS}$ scheme by \cite {MulTer}\footnote{In ref. 
\cite {MulTer} these transformations are presented in Bjorken x space
to all orders in $\alpha_s$.}:
\begin{equation}
\Delta C^{(1)}_q(n)_{JET}=\Delta C^{(1)}_q(n)_{\overline{MS}}~,~~~~
\Delta C^{(1)}_G(n)_{JET}=\Delta C^{(1)}_G(n)_{\overline{MS}}
- {2N_f\over n(n+1)}~,
\label{Ctransf}
\end{equation}
\begin{eqnarray}
\nonumber
\Delta \gamma^{(1)}_{qq}(n)_{JET}&=&
\Delta \gamma^{(1)}_{qq}(n)_{\overline{MS}}+{4N_f\over n(n+1)}
\Delta \gamma^{(0)}_{Gq}(n)~,\\
\nonumber
\Delta \gamma^{(1)}_{qG}(n)_{JET}&=&
\Delta \gamma^{(1)}_{qG}(n)_{\overline{MS}}+{4N_f\over n(n+1)}
[\Delta \gamma^{(0)}_{GG}(n) - \Delta \gamma^{(0)}_{qq}(n)
+2\beta_0]~,\\
\nonumber
\Delta \gamma^{(1)}_{Gq}(n)_{JET}&=&
\Delta \gamma^{(1)}_{Gq}(n)_{\overline{MS}}~,\\
\Delta \gamma^{(1)}_{GG}(n)_{JET}&=&
\Delta \gamma^{(1)}_{GG}(n)_{\overline{MS}}-{4N_f\over n(n+1)}
\Delta \gamma^{(0)}_{Gq}(n)~.
\label{anomdimtransf}
\end{eqnarray}

Note also that
\begin{equation}
\Delta C^{(1)}_{NS}(n)_{JET}=
\Delta C^{(1)}_{NS}(n)_{\overline{MS}}~,~~~~~~~~
\Delta \gamma^{(1)}_{NS}(n)_{JET}=
\Delta \gamma^{(1)}_{NS}(n)_{\overline{MS}}~.
\label{NStransf}
\end{equation}
In (\ref {anomdimtransf}) a superscript "0"  is used for the
corresponding anomalous dimensions in the LO approximation.

Note that the transformation of the coefficient functions and
anomalous dimensions from the $\overline{MS}$ to the AB scheme is
given by eqs. (\ref{Ctransf} - \ref{NStransf}), in which the factor
$2/n(n+1)$ should be replaced by $1/n$.\\
 
In each scheme the parton densities at $Q^2_0=1~GeV^2$, as in
\cite{LSiSt98}, were parametrized in the form:
\begin{eqnarray}
\nonumber
x\Delta u_v(x,Q^2_0)&=&\eta_u A_ux^{a_u}xu_v(x,Q^2_0)~,\\
\nonumber
x\Delta d_v(x,Q^2_0)&=&\eta_d A_dx^{a_d}xd_v(x,Q^2_0)~,\\
\nonumber
x\Delta Sea(x,Q^2_0)&=&\eta_S A_Sx^{a_S}xSea(x,Q^2_0)~,\\
x\Delta G(x,Q^2_0)&=&\eta_g A_gx^{a_g}xG(x,Q^2_0)
\label{classic}
\end{eqnarray}
where on R.H.S. of (\ref{classic}) we have used the recent MRST 
unpolarized densities \cite{MRST}. The normalization factors 
$~A_f~$ are determined in such a way as to ensure that the first 
moments of the polarized densities are given by $~\eta_{f}$.

The first moments of the valence quark  densities
$~\eta_u~$ and $~\eta_d~$ are fixed by the octet nucleon and 
hyperon $\beta$ decay constants \cite{PDG}
\begin{equation}
g_{A}=F+D=1.2573~\pm~0.0028,~~~a_8=3F-D=0.579~\pm~0.025~.
\label{GA3FD}
\end{equation}
and in the case of SU(3) flavour symmetry of the sea  
($\Delta\bar{u}=\Delta\bar{d}=\Delta\bar{s}~$ at $~Q^2_0$)
\begin{equation}
\eta_u=0.918~,~~~~~~~~\eta_d=-0.339~.
\label{etaudSU3}
\end{equation}

The rest of the parameters in (\ref{classic})
\begin{equation}
\{a_u,~a_d,~\eta_S,~a_S~,\eta_g~,a_g\}~,
\label{claspar}
\end{equation}
have to be determined from the best fit to the $~A_1^N(x,Q^2)~$ data.

To calculate the virtual photon-nucleon asymmetry
$~A_1^N(x,Q^2)~$ in NLO QCD and then fit to the data we follow
the procedure described in detail in our previous papers \cite{LSiSt,
LSiSt98}, where the connection between measured quantities,
Wilson coefficients and parton densities is given.\\

\def\thefootnote{\dagger}
The numerical results of the fits in the JET, AB and $\overline{MS}$ 
schemes to the present experimental data on $~A_1^N(x,Q^2)~$
\cite{E154, newSMCpQ2, finSLACpd, EMC, SLACnQ2, SMCd97Q2, HERMES}  
are listed in Table 1.\footnote{After the completion of this work
new data on $A^p_1$ and $g^p_1$ have been reported by the HERMES 
Collaboration \cite{HERMESp}.} 
The data used (118 experimental points) cover the following 
kinematic region:  
\begin{equation}
0.004< x < 0.75,~~~~~~1< Q^2< 72~GeV^2~.
\label{kinreg}
\end{equation}

In this paper we present the results of the fit to the $A^N_1$ data
averaged over $Q^2$ at each $x$. (In our previous work \cite{LSiSt98})
we also analyzed data in $(x,Q^2)$ bins. Our conclusion in the present
paper also hold for this type of fit.) The total (statistical and 
systematic) errors
are taken into account. The results presented in Table 1
correspond to an SU(3) symmetric sea. Note that in this case the
first moment of the strange sea quarks $\eta_{\bar{s}}\equiv
\Delta\bar{s}(1,Q^2_0)=\eta_S/6~$. As was shown in our previous
work \cite{LSiSt98}, the flavour decomposition of the sea does
not affect the quality of the fit and the results on the
polarized parton densities $\Delta \Sigma$, $\Delta\bar{s}$ and 
$\Delta G$.

As in the case of the $\overline{MS}$ scheme \cite{LSiSt98}, the 
value of $a_g$ is not well determined by the fits in the JET 
and AB schemes to the existing data , i.e. $~\chi^2/DOF~$ 
practically does not change when $a_g$ varies in the 
range:$0\leq a_g\leq 1$. In Table 1 we present the results of 
the fits corresponding to $a_g=0.6$.\\
\vskip 0.6 cm
\begin{center}
\begin{tabular}{cl}
&{\bf Table 1.} Results of the NLO QCD fits in the JET, AB and 
$\overline{MS}$ schemes to\\ 
&the world $~A_1^N~$ data ($Q^2_0=1~GeV^2$). The errors shown 
are total (statistical\\ 
&and systematic). $a_g=0.6$ (fixed).
\end{tabular}
\vskip 0.6 cm
\begin{tabular}{|c|c|c|c|c|c|c|} \hline
 ~~Scheme~~&~~~~~~~~~~~JET~~~~~~~~~~~&~~~~~~~~~~~AB~~~~~~~~~~~
 &~~~~~~~~~~~$\overline{MS}$~~~~~~~~~~~\\ \hline
 $DOF$    &  118~-~5    &     118~-~5   &   118~-~5\\
 $\chi^2$        &  86.39    &   86.35  &   86.11  \\
 $\chi^2/DOF$    &  0.764    &   0.764  &   0.762   \\  \hline
 $a_u$           &~~0.267~~$\pm$~~0.035~~ &~~0.267~~$\pm$~~0.036~~
 &~~0.255~~$\pm$~~0.028 \\
 $a_d$           &~~0.124~~$\pm$~~0.123~~&~~0.124~~$\pm$~~0.125~~
 &~~0.148~~$\pm$~~0.113\\
 $a_S$           &~~1.469~~$\pm$~~0.460~~&~~1.558~~$\pm$~~0.583~~
 &~~0.817~~$\pm$~~0.223\\
 $\eta_{\bar{s}}$&-~0.027~~$\pm$~~0.004~~&-~0.022~~$\pm$~~0.005~~ 
 &-~0.049~~$\pm$~~0.005\\
 $\eta_g$        &~~~0.50~~$\pm$~~0.12~~&~~0.56~~$\pm$~~0.14~~ 
 &~~~0.82~~$\pm$~~0.32~~\\ \hline
 $\Delta \Sigma(1)$&~~0.416~~$\pm$~~0.036~~&~~0.444~~$\pm$~~0.040~~
&~~0.287~~$\pm$~~0.041\\ 
 $a_0(1~GeV^2)$&~~0.30~~$\pm$~~0.04&~~0.31~~$\pm$~~0.05~~
 &~~0.29~~$\pm$~~0.04\\  \hline
\end{tabular}
\end{center}
\vskip 0.6 cm

It is seen from the Table 1 that the values of $~\chi^2/DOF~$ 
coincide almost exactly in the different factorization schemes, 
which is a good indication of the stability of the analysis.
The NLO QCD predictions are in a very good
agreement with the presently available data on $A^N_1$, as is
illustrated in the JET scheme fit in Fig. 1.
We would like to draw special attention to the excellent fit to the
very accurate E154 neutron data 
($\chi^2=1.1$ for 11 experimental data points).

The extracted valence and gluon polarized densities at $Q^2_0=1~GeV^2$
for the different schemes are shown in Fig. 2 (solid, dotted and
dashed curves correspond to JET, AB and $\overline{MS}$ scheme,
respectively). 
Note that the valence densities $x\Delta u_v$ and $x\Delta d_v$
in the JET and AB schemes are almost identical so the dotted curves 
corresponding to $x\Delta u_v$ and $x\Delta d_v$
are not shown in Fig. 2. The difference between $x\Delta u_v$ and 
$x\Delta d_v$ in the $\overline{MS}$ and JET scheme is negligible. 
The results of the fit for the polarized valence densities are in
an excellent agreement with what follows from the theory, namely, 
that they should be the same in the factorization schemes under
consideration.

The results on the polarized gluon densities determined by the
fit in the JET, AB and $\overline{MS}$ schemes are also
consistent. The values of their first moments $\eta_g$ coincide 
within errors (see Table 1). However, as a
consequence of the uncertainty in determining the gluon density
from the present data, the central values of $\eta_g$ and
therefore, the gluon densities themselves, differ somewhat in
the various schemes.\\

In Table 1 we also present our results in the different schemes
for the first moments $\Delta \Sigma(1)$ of the polarized singlet
quark density as well as for the axial charge $a_0$. The obtained 
singlet densities $\Delta \Sigma(x,Q^2_0)$ are shown in Fig 3a. 
It is seen from the table that the first moments 
$\Delta \Sigma(1)$ in the JET and AB
schemes are in a very good agreement. (We recall that according
to the definition of the JET and AB schemes $\Delta \Sigma(1)$
should be the same in the both schemes.) The corresponding densities
$\Delta \Sigma(x,Q^2)$, however, are slightly different (see Fig.
3a) because their higher moments are not equal.

Our result for $\Delta \Sigma(1)_{\overline{MS}\Rightarrow JET,AB}$
using the values of $\Delta \Sigma(1)_{\overline{MS}}$ and 
$(\eta_g)_{\overline{MS}}$ from Table 1 and the NLO transformation
rules (see eqs. (\ref{ABMSmom}) and (\ref{JETMS}) for $n=1$) 
\begin{equation}
\Delta \Sigma(1)_{\overline{MS}\Rightarrow JET,AB}=0.476\pm 0.084
\label{sig1ABour}
\end{equation}
coincides within errors with the values of $\Delta \Sigma(1)_{JET}$ 
and $\Delta \Sigma(1)_{AB}$
\begin{equation}
\Delta \Sigma(1)_{JET}=0.416\pm 0.036,~~~\Delta \Sigma(1)_{AB}=
0.444\pm 0.040
\label{sig1ABour}
\end{equation}
determined directly by the fit to the data in the JET and AB
schemes as presented in Table 1. The singlet densities 
$\Delta \Sigma(x,Q^2_0)_{JET}$ (solid curve) and $\Delta \Sigma
(x,Q^2_0)_{\overline{MS}\Rightarrow JET}$ (dashed curve) are
shown in Fig. 3b.

Finally, we would like to draw attention to the excellent
agreement between the values of the axial charge $a_0(1~GeV^2)$
determined in the different schemes (see Table 1), which  
illustrates impressively how our analysis respects the 
scheme-independence of physical quantities.\\  

In conclusion, we have performed a next-to leading order QCD 
analysis of the world data on inclusive 
polarized deep inelastic lepton-nucleon scattering in the JET, 
AB and $\overline{MS}$ schemes.
The QCD predictions have been confronted with the data on the
virtual photon-nucleon asymmetry $~A_1^N(x,Q^2)~$, rather than
with the polarized structure function $~g_1^N(x,Q^2)~$, in order
to minimize higher twist effects. Using the simple 
parametrization (\ref{classic}) (with only 5 free parameters) 
for the input polarized parton densities it was
demonstrated that the polarized DIS data are in an excellent
agreement with the pQCD predictions for $~A_1^N(x,Q^2)~$ in all 
the factorization schemes considered

Moreover, we have demonstrated the consistency between the results
of the analysis in each scheme and the NLO transformation
equations relating them. Such consistency gives us confidence
that the NLO analysis is reliable.\\

{\bf Acknowledgments}
\vskip 3mm

We are grateful to O. V. Teryaev for useful discussions and remarks.\\

This research was partly supported by a UK Royal Society Collaborative
Grant, by the Russian Fund for Fundamental Research Grant No 
96-02-17435a and by the Bulgarian 
Science Foundation under Contract \mbox{Ph 510.}\\


\newpage
\noindent
{{\bf Figure Captions}}
\vskip 3mm
\noindent 
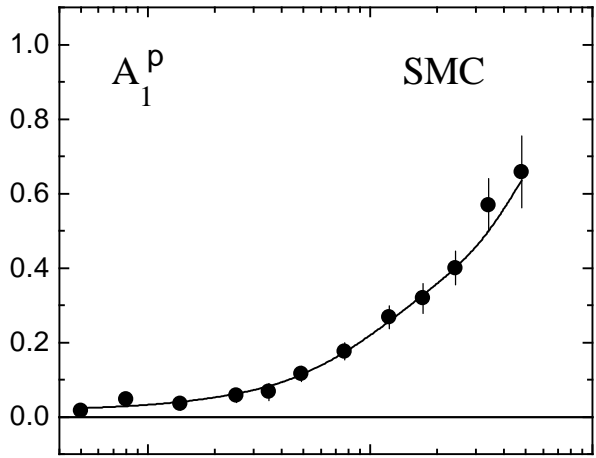
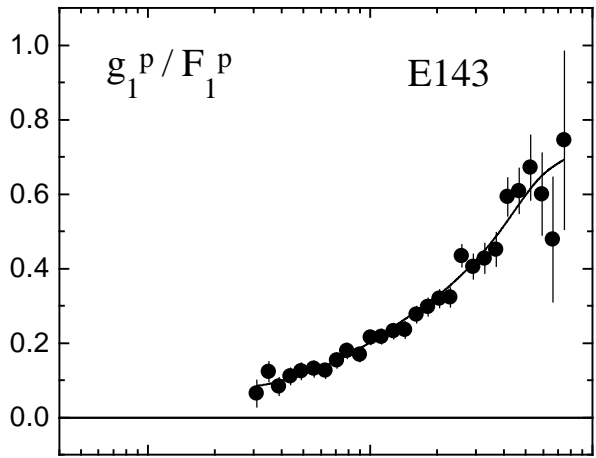
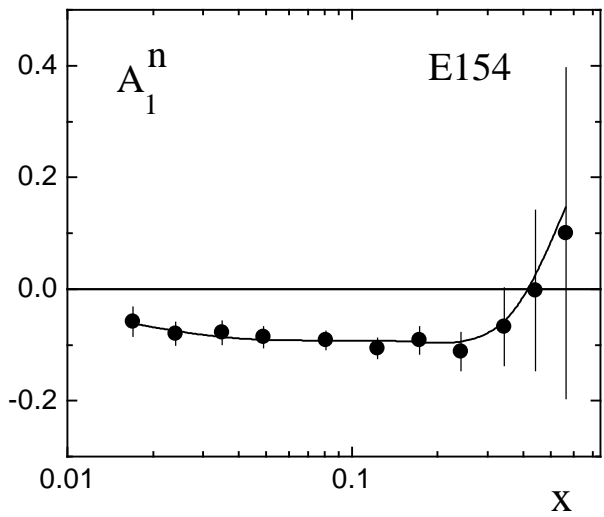
{\bf Fig. 1} Comparison of our NLO results in the JET scheme
for $~A_1^N(x,Q^2)~$ with the experimental data at the 
measured $x$ and $Q^2$ values. Errors bars represent the 
total error.\\

\noindent
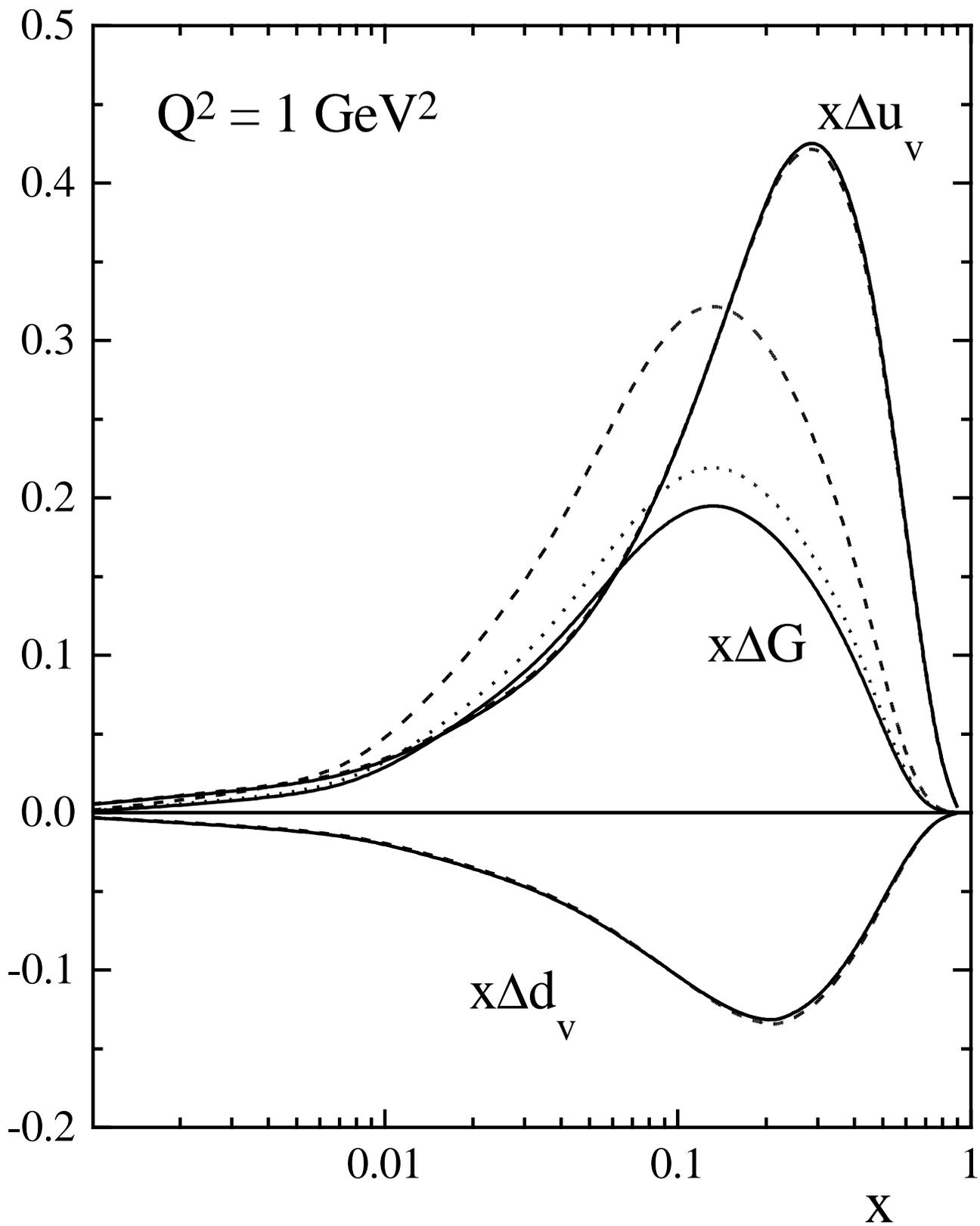
{\bf Fig. 2} Next-to-leading order input polarized valence and gluon
distributions at $~Q^2=1~GeV^2~$ in different factorization schemes. 
Solid, dotted and dashed curves correspond to the JET, AB and 
$\overline{MS}$ scheme, respectively.\\

\noindent
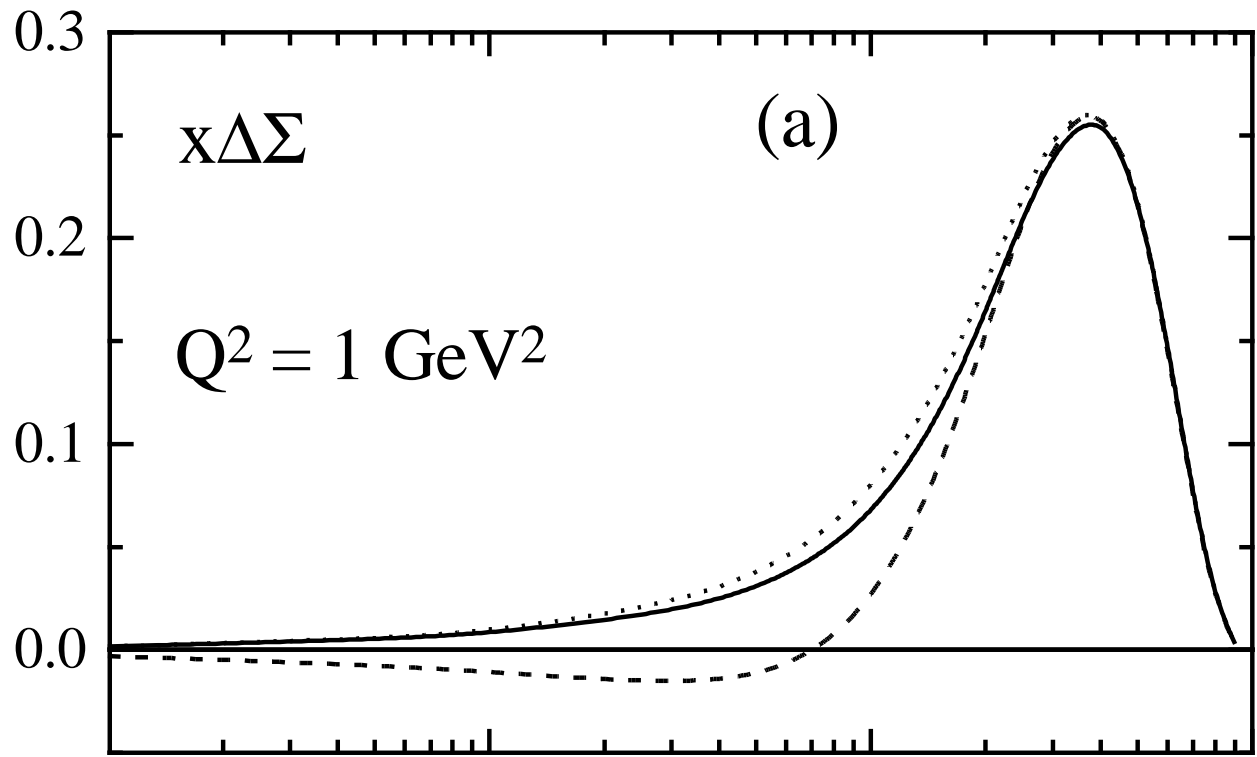
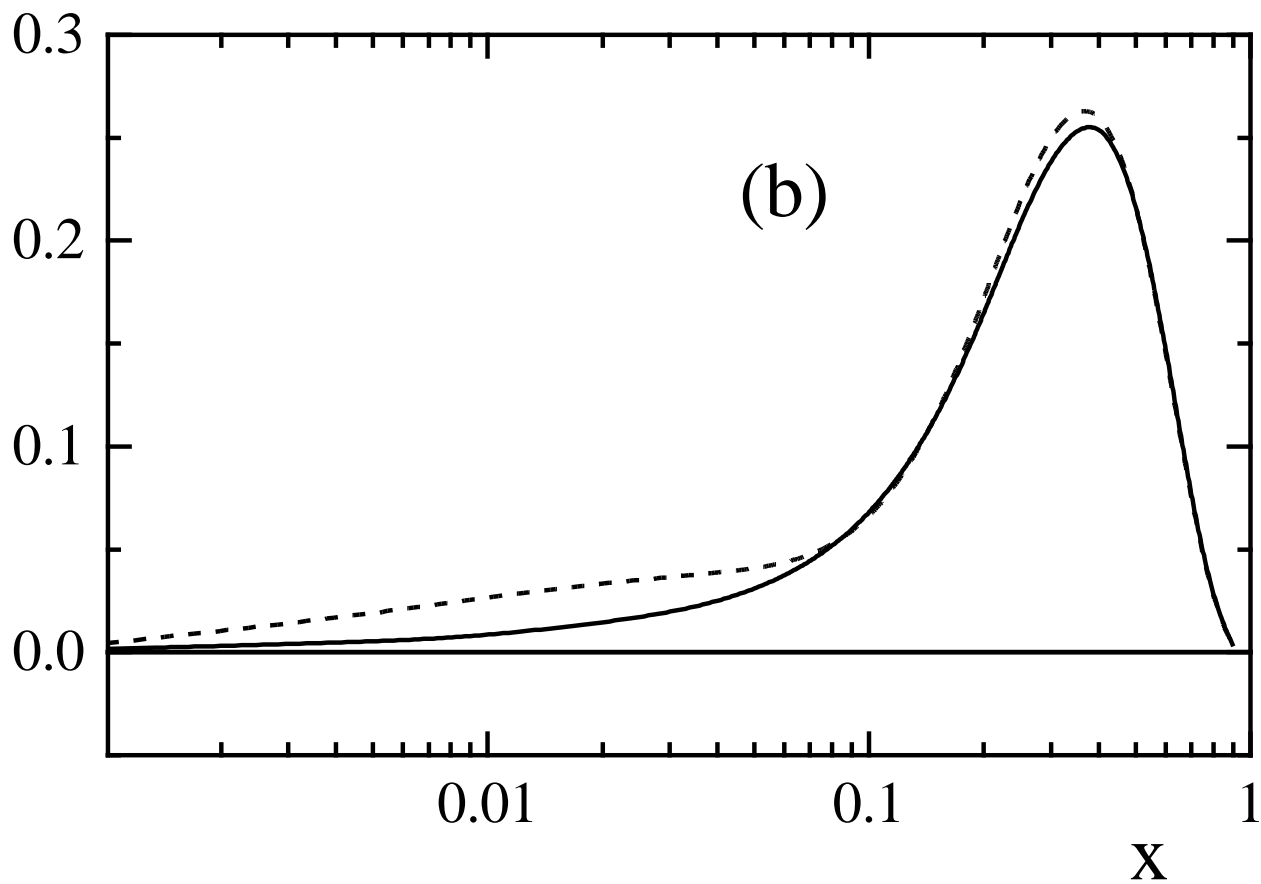
{\bf Fig. 3} (a) Next-to-leading order input polarized singlet
distributions $\Delta \Sigma(x)$ at $~Q^2=1~GeV^2~$ in different 
factorization schemes. Solid, dotted and dashed curves correspond 
to the JET, AB and $\overline{MS}$ scheme, respectively. 
(b) Comparison between the singlet density $\Delta \Sigma(x)_{JET}$
obtained from the fit (solid curve)
and $\Delta \Sigma(x)_{\overline{MS}\Rightarrow JET}$ determined 
by eq. (\ref{JETMS}) (dashed curve), at $~Q^2=1~GeV^2~$.

\end{document}